\documentclass[pdflatex,sn-mathphys-num]{sn-jnl}
\usepackage[dvipsnames,table]{xcolor} 
\usepackage{graphicx}%
\usepackage{multirow}%
\usepackage{amsmath,amssymb,amsfonts}%
\usepackage{amsthm}%
\usepackage{mathrsfs}%
\usepackage[title]{appendix}%
\usepackage{textcomp}%
\usepackage{manyfoot}%
\usepackage{booktabs}%
\usepackage{algorithm}%
\usepackage{algorithmicx}%
\usepackage{algpseudocode}%
\usepackage{listings}%
\usepackage{amsmath}
\usepackage{balance}
\usepackage{graphicx}
\usepackage{enumitem}
\usepackage{booktabs}
\usepackage{subcaption}
\usepackage{multirow}
\usepackage{float}
\usepackage{xcolor}
\usepackage{float}    
\usepackage{array}    

\renewcommand\fbox{\fcolorbox{blue}{white}}
\usepackage{adjustbox}

\usepackage{xspace}
\usepackage{xcolor}
\usepackage{url}
\usepackage{makecell} 
\usepackage{array}    
\usepackage{colortbl}
\usepackage{arydshln}
\usepackage[most]{tcolorbox}
\usepackage[table]{xcolor}
\usepackage{multirow}
\usepackage{booktabs}
\usepackage{colortbl}
\usepackage{adjustbox}

\newcommand{\llm}[0]{LLM\xspace}
\newcommand{\llms}[0]{LLMs\xspace}

\newcommand{\slms}[0]{SLMs\xspace}
\newcommand{\task}[0]{SVD\xspace}

\newcommand{\codeqwen}[0]{CodeQwen1.5\xspace}
\newcommand{\deepseek}[0]{DeepSeek-Coder\xspace}
\newcommand{\codegemma}[0]{CodeGemma\xspace}
\newcommand{\starcoder}[0]{Starcoder-2\xspace}
\newcommand{\codellama}[0]{CodeLlama\xspace}

\theoremstyle{thmstyleone}%
\theoremstyle{thmstyletwo}%

\theoremstyle{thmstylethree}%

\begin{document}

\title[Article Title]{Benchmarking Large Language Models for Multi-Language Software Vulnerability Detection}

\author{Ting Zhang\textsuperscript{1,*}, Chengran Yang\textsuperscript{1,*}, Yindu Su\textsuperscript{1}, Martin Weyssow\textsuperscript{1}, Hung Nguyen\textsuperscript{1}, Tan Bui\textsuperscript{1}, Hong Jin Kang\textsuperscript{2}, Yikun Li\textsuperscript{1}, Eng Lieh Ouh\textsuperscript{1}, Lwin Khin Shar\textsuperscript{1}, David Lo\textsuperscript{1}}

\affil{
  \textsuperscript{1}School of Computing and Information Systems, Singapore Management University, Singapore \\
  Email: \{tingzhang.2019, cryang, yindusu, mweyssow, huuhungn, ngoctanbui, yikunli, elouh, lkshar, davidlo\}@smu.edu.sg \\
  \textsuperscript{2}School of Computer Science, University of Sydney, Australia \\
  Email: hongjin.kang@sydney.edu.au
}

\affil[*]{Both authors contributed equally to this research}

\abstract{Recent advancements in generative AI have led to the widespread adoption of large language models (LLMs) in software engineering, addressing numerous longstanding challenges. However, a comprehensive study examining the capabilities of LLMs in software vulnerability detection (SVD), a crucial aspect of software security, is currently lacking.
Existing research primarily focuses on evaluating LLMs using C/C++ datasets. 
It typically explores only one or two strategies among prompt engineering, instruction tuning, and sequence classification fine-tuning for open-source LLMs. 
Consequently, there is a significant knowledge gap regarding the effectiveness of diverse LLMs in detecting vulnerabilities across various programming languages.

To address this knowledge gap, we present a comprehensive empirical study evaluating the performance of LLMs on the SVD task. 
We have compiled a comprehensive dataset comprising 8,260 vulnerable functions in Python, 7,505 in Java, and 28,983 in JavaScript.
We assess five open-source LLMs using multiple approaches, including prompt engineering, instruction tuning, and sequence classification fine-tuning.
These LLMs are benchmarked against five fine-tuned small language models and two open-source static application security testing tools.
Furthermore, we explore two avenues to improve LLM performance on SVD: a) Data perspective: Retraining models using downsampled balanced datasets. b) Model perspective: Investigating ensemble learning methods that combine predictions from multiple LLMs.

Our comprehensive experiments demonstrate that SVD remains a challenging task for LLMs. This study provides a thorough understanding of the role of LLMs in SVD and offers practical insights for future advancements in leveraging generative AI to enhance software security practices.
}


\maketitle

\section{Introduction}
Vulnerabilities pose a significant threat to software security.
Many cybersecurity incidents and data breaches are caused by exploitable software vulnerabilities (SVs)~\cite{liu2018detecting,lin2019software}. 
Consequently, automatic software vulnerability detection (SVD) has become a crucial task.
In the past decade, numerous static analysis or dynamic analysis tools have been proposed to address this challenge~\cite{cadar2008klee,Flawfind20:online,Applicat17:online}. However, both types of tools have drawbacks: static tools require substantial manual effort from security experts to craft rules, and they have limited generalization ability across diverse vulnerabilities; dynamic tools require configuring execution, which is usually complex, and execution results may be incomplete since not every program path can be executed~\cite{liu2024enhancing}.

Given that deep learning requires no manual operations to find rules and does not need extensive feature engineering, the recent trend in automatic SVD is to apply advancements made in deep learning~\cite{li2021sysevr,li2018vuldeepecker,lin2019software}. 
For instance, the first deep-learning-based approach for SVD, VulDeePecker~\cite{li2018vuldeepecker}, adopts bi-directional Long Short-Term Memory~\cite{cho2014properties}. 
While more recent endeavors have started to utilize pre-trained Transformer models, such as LineVul~\cite{fu2022linevul}, these pre-trained Transformer models were prevalent before the era of large language models (LLMs) began.

LLMs, such as GPT-4~\cite{gpt4:online} and Claude~\cite{claude:online}, have revolutionized natural language understanding and generation. In recent years, LLMs have been rapidly evolving and bringing considerable breakthroughs in the field of natural language processing. Moreover, these LLMs have also begun to impact various software engineering tasks, e.g., sentiment analysis in software engineering~\cite{zhang2023revisiting}, automatic program repair~\cite{xia2023keep}, and duplicate bug report detection~\cite{zhang2023cupid}. However, the effectiveness of LLMs for SVD remains unclear. Recently, a few empirical studies have attempted to evaluate the effectiveness of LLMs for SVD~\cite{ding2024vulnerability,zhang2024empirical,steenhoek2024comprehensive,shestov2024finetuning,gao2023far}. We identify the following limitations in the existing empirical studies.

\vspace{4px}
\noindent{\textbf{Focusing on C/C++ vulnerabilities:}} Most studies evaluating LLMs on real-world vulnerability data focus primarily on C/C++ vulnerabilities~\cite{shestov2024finetuning,steenhoek2024comprehensive,gao2023far}. 
However, there is a notable gap in research concerning vulnerabilities in other programming languages (PLs).
Given the difference in the language design and features, the vulnerabilities in different PLs would also differ.
For instance, C/C++ lacks automatic bounds checking, making it susceptible to buffer overflow vulnerabilities. 
In contrast, other PLs, such as Python, Java, and JavaScript, implement automatic bounds checking for arrays and other data structures, substantially reducing the risk of buffer overflows.
Notably, Python, Java, and JavaScript are the top-3 most popular PLs in GitHub open-source projects from 2015 to 2022.\footnote{\url{https://github.blog/news-insights/research/the-state-of-open-source-and-ai/\#the-most-popular-programming-languages}}
As two of the most popular and well-developed PLs, Python and Java have a broad scope of application scenarios.
Unfortunately, recent empirical studies evaluating LLMs have not adequately addressed vulnerabilities in these and other PLs beyond C/C++.
While there are several datasets containing Python or Java vulnerabilities in the literature, they are either more coarse-grained or contain few vulnerabilities, e.g., CrossVul~\cite{nikitopoulos2021crossvul}, which only provide data at the file level, not the function level; Vul4J~\cite{bui2022vul4j}, which contains only 79 vulnerabilities from 25 CWE types.

\vspace{4px}
\noindent{\textbf{Lack of comprehensive evaluation of open-source LLMs:}} Existing studies typically conduct only one or two strategies among prompt engineering, instruction tuning, and sequence classification fine-tuning on open-source LLMs. 
However, a comprehensive comparison of all three approaches is lacking. 
Such a comparison would provide a deeper understanding of how each method impacts the effectiveness of LLMs across different PLs in the SVD task.
For instance, Guo et al.~\cite{gao2023far} only conducted few-shot learning on LLMs while not tuning the parameters of LLMs; 
similarly, Steenhoek et al.~\cite{steenhoek2024comprehensive} conducted in-context learning with multiple prompt engineering methods on LLMs without tuning any parameters. 
Ding et al.~\cite{ding2024vulnerability} only conducted sequence classification fine-tuning on open-source LLMs, while they did not conduct few-shot prompting or instruction-tuning on open-source LLMs.

Additionally, all the existing studies have not compared LLMs with their smaller counterparts\footnote{We refer to the LLMs with less than 1B parameters as SLMs, which stands for ``small language models''.} and static application security testing (SAST) tools together.
On the one hand, a prior study on sentiment analysis for software engineering shows that SLMs outperform LLMs when provided with sufficient training data~\cite{zhang2023revisiting}. However, it remains unclear whether this finding extends to SVD.
On the other hand, while SAST tools are widely used in practice for identifying vulnerabilities~\cite{li2023comparison}, their effectiveness relative to LLMs in this domain has not been thoroughly evaluated.

To bridge this gap, in this study, we aim to conduct a more comprehensive empirical study to answer the following research questions:

\begin{itemize}[leftmargin=*]
\item \textbf{RQ1:} \emph{To what extent are LLMs effective in predicting SVs?}
\item \textbf{RQ2:} \emph{How does the effectiveness of LLMs in predicting SVs compare to that of SLMs and SAST tools?}
\item \textbf{RQ3:} \emph{What strategies can be employed to enhance the effectiveness of LLMs in predicting SVs?}
\end{itemize}

Our contributions {\bf in the context of SVD} can be summarized as follows:

\begin{itemize}[leftmargin=*]
\item \textbf{Comprehensive investigation on LLMs:} We evaluate different usages of five open-source LLMs, from freezing parameters (prompt engineering) to tuning parameters (instruction tuning and sequence classification fine-tuning).
\item \textbf{Comparative study of LLMs with SLMs and SAST tools:} We compare LLMs with four SLMs and two SAST tools.
\item \textbf{Attempts to improve LLMs:} We also investigate whether (1) training LLMs with downsampled data and (2) combining multiple LLMs and SLMs can be helpful in improving the accuracy of the SVD task. 
\item \textbf{Dataset:} We provide a dataset comprised of vulnerability data from projects written in the top three most popular PLs, i.e., Python, Java, and JavaScript.
\end{itemize}

The structure of this paper is as follows.
Section~\ref{sec:related_work} discusses the background and related work.
Section~\ref{sec:overview} presents the overview of our study details of the investigated LLMs.
We describe the experimental design in Section~\ref{sec:setup}.
Section~\ref{sec:results} presents the experimental results.
We discuss the implications and threats to validity in Section~\ref{sec:discussion}.
Finally, Section~\ref{sec:conclusion} concludes this paper and discusses future work.

\section{Background and Related Work}
\label{sec:related_work}
In recent years, vulnerability research has gained significant attention.
One research line is proposing deep-learning-based approaches for SVD ~\cite{fu2022linevul,zhang2023vulnerability,lu2024grace,cao2024coca};
another research line is conducting empirical studies to evaluate the effectiveness of various models and tools~\cite{steenhoek2023empirical,steenhoek2024comprehensive,zhang2024empirical,ding2024vulnerability,li2023comparison,zhou2024large}.
In this section, we briefly review relevant studies in two research directions: leveraging LMs for SVD and evaluating LMs for SVD.
For the purposes of this study, we use the term ``Language Models" (LMs) to encompass both SLMs and LLMs.

\subsection{Leveraging LMs for SVD}
SLMs have been widely used for SVD, as they are relatively small and require fewer computational resources.
For instance, Grace~\cite{lu2024grace} is a vulnerability detection approach based on LLM, and it leverages the CodeT5~\cite{wang2021codet5} model to extract code semantic features.
Vulberta~\cite{hanif2022vulberta} pre-trains a RoBERTa model with a custom tokenization pipeline that has been leveraged to train vulnerability detection classifiers.
Fu et al.~\cite{fu2022linevul} proposed LineVul, a Transformer-based line-level vulnerability prediction approach that leverages the attention mechanism within the BERT architecture for line-level predictions. Through an empirical evaluation on a large-scale real-world dataset with 188k+ C/C++ functions, LineVul achieved (1) 160\%-379\% higher F1-score for function-level predictions and (2) 12\%-25\% higher Top-10 Accuracy for line-level predictions compared to baselines.

Zhang et al.~\cite{zhang2023vulnerability} introduced EPVD, which decomposes code snippets into execution paths for SVD. 
EPVD utilizes the pre-trained CodeBERT model to learn intra-path attentions and encodes each path into a feature vector. Evaluated on over 231K functions from three high-quality C/C++ datasets, EPVD significantly outperformed state-of-the-art approaches.

The aforementioned works were evaluated on C/C++ datasets, and the largest LM considered was GPT-J~\cite{gao2020pile} with 6B parameters. 
In contrast, our work utilizes a newly curated dataset with Python, Java, and JavaScript and includes a comparison of LLMs with up to 8.5B parameters.
In addition, our study incorporates the recent popular top-performing open-source LLMs, e.g., \codeqwen and \deepseek.

\subsection{Evaluating LMs for SVD}
Several studies have compared the performance of fine-tuned SLMs for SVD~\cite{yin2024multitask,liu2024vuldetectbench}. 
Thapa et al.~\cite{thapa2022transformer} conducted an empirical analysis of C/C++ source codes with multiple vulnerabilities related to library function calls, pointer usage, array usage, and arithmetic expressions. Their results demonstrated the good performance of LLMs in vulnerability detection and their superiority over traditional models like bidirectional LSTMs and GRUs in terms of F1-score.

Chen et al.~\cite{chen2023diversevul} proposed DiverseVul, a dataset containing 18,945 vulnerable functions spanning 155 CWEs and 330,492 non-vulnerable functions extracted from 7,514 commits. On this dataset, they experimented with 11 deep learning architectures from four model families: Graph Neural Networks, RoBERTa, GPT-2, and T5. Their results showed that LLMs outperformed state-of-the-art graph neural networks for deep learning-based SVD.

The two most relevant empirical studies were conducted by Ding et al.~\cite{ding2024vulnerability} and Steenhoek et al.~\cite{steenhoek2024comprehensive}. 
Ding et al. evaluated 7 code LLMs of varying sizes, including state-of-the-art open-source models like StarCoder 2~\cite{lozhkov2024starcoder} and proprietary models from OpenAI~\cite{achiam2023gpt}, on their proposed PrimeVul dataset. 
They fine-tuned LLMs with fewer than 10B parameters and conducted few-shot prompting and chain-of-thought for larger LLMs. 
Their results found that code LLMs consistently performed poorly on the PrimeVul benchmark.

Steenhoek et al.~\cite{steenhoek2024comprehensive} evaluated 11 state-of-the-art LLMs commonly used as coding assistants for their capabilities in SVD. They systematically searched for the best-performing prompts, incorporating techniques such as in-context learning and chain-of-thought, and proposed three prompting methods. Their results showed that LLMs generally struggled with SVD tasks.

In summary, our work distinguishes itself from existing studies in terms of (1) dataset diversity: We evaluate the techniques on real-world vulnerabilities across various PLs; (2) tool selection: We evaluate LLMs, SLMs, and SAST tools; and (3) LLM implementation: We explore multiple strategies for LLM adoption, including prompt engineering (where all parameters are frozen) and instruction tuning and sequence classification fine-tuning (where parameters are tuned).

\section{Benchmark and LMs Studied}
\label{sec:overview}

\subsection{Benchmark Creation}

\begin{table*}[ht]
    \centering
    \caption{Data statistics for the time-aware split.}
    \begin{adjustbox}{max width=\textwidth} 
    \begin{tabular}{c c c c c c c c c c}
    \toprule
        \multirow{3}{*}{PL} &   \multicolumn{3}{c}{Training} &  \multicolumn{3}{c}{Validation} &  \multicolumn{3}{c}{Testing}  \\
        \cmidrule{2-10}
       & \multirow{2}{*}{\# VFCs} & \multicolumn{2}{c}{Functions} & \multirow{2}{*}{\# VFCs} & \multicolumn{2}{c}{Functions} &  \multirow{2}{*}{\# VFCs} &  \multicolumn{2}{c}{Functions} \\
       \cmidrule{3-4}
       \cmidrule{6-7}
       \cmidrule{9-10}
       & & \# Vuln. & \# Non-vuln. &  & \# Vuln. & \# Non-vuln. & & \# Vuln. & \# Non-vuln.\\
        \midrule
        \textbf{Python} & 1,014 & 4,465 & 71,649 & 33 & 1,758 & 6,705 & 247 & 2,037 & 21,108 \\
        \textbf{Java} & 871 & 6,354 & 63,654 & 80 & 616 & 7,209 & 99 & 535 & 6,600 \\
        \textbf{JavaScript} & 1,029 & 22,198 & 42,409 & 78 & 3,418 & 4,053 & 109 & 3,367 & 15,862 \\
    \bottomrule
    \end{tabular}  
    \end{adjustbox} 
    \label{tab:dataset}
\end{table*}

\begin{figure}
    \centering
    \includegraphics[width=0.9\textwidth]{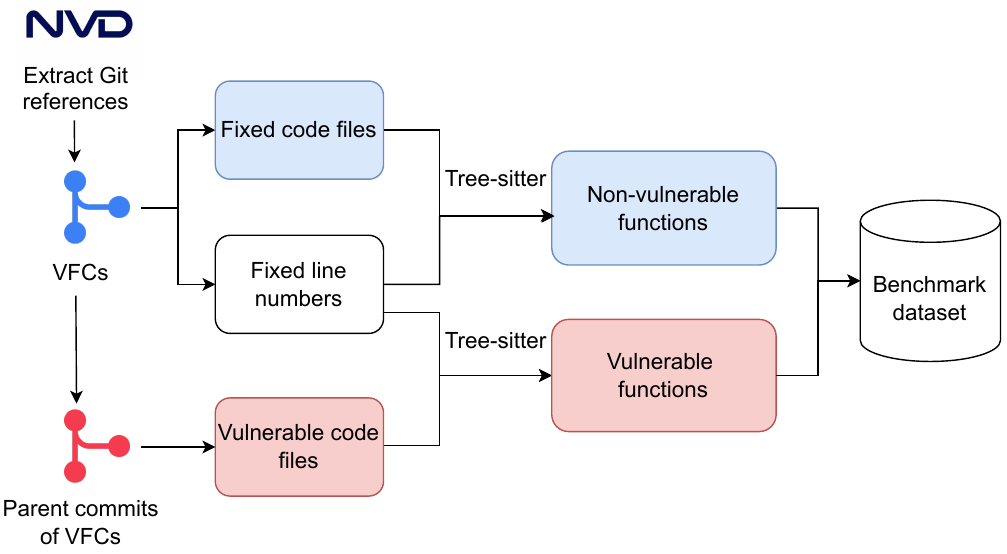}
    \caption{Pipeline for benchmark crafting.}
    \label{fig:benchmark}
\end{figure}

This section describes the vulnerability dataset we mined for this task.
Based on the IEEE Spectrum survey\footnote{\url{https://spectrum.ieee.org/top-programming-languages-2024}}, the top three PLs in 2023 were Python, Java, and JavaScript. 
Therefore, we focus on vulnerabilities in these three PLs. 

Figure~\ref{fig:benchmark} illustrates the detailed process of extracting vulnerable functions.
We collect vulnerability data from the National Vulnerability Database (NVD), a comprehensive repository of publicly disclosed cybersecurity vulnerabilities, up to March 28, 2024.
Specifically, NVD maintains vulnerabilities as a list of Common Vulnerabilities and Exposures (CVEs). 
For each CVE, we search for Vulnerability-Fixing Commits (VFCs) by identifying references in the CVE entry that include links containing a ``patches'' tag.
We particularly look for links from GitHub, GitLab, or BitBucket repositories.
We consider VFCs from $\langle PL\rangle$, where $\langle PL \rangle \in \{Python, Java, JavaScript\}$, if they modify at least one file written in the corresponding language.
Following prior works~\cite{chakraborty2021deep,chen2023diversevul,fan2020ac}, which assume VFCs only modify vulnerable functions to address security flaws,
we label the pre-commit versions of changed functions as \textit{vulnerable}, while labeling the post-commit versions and all unchanged functions as \textit{non-vulnerable}.
We apply Tree-sitter~\cite{treesitter:online} to extract changed and unchanged functions from VFCs.
Following previous studies~\cite{chen2023diversevul}, we perform function de-duplication to prevent data leakage by using MD5 hashes without applying code normalization.
We maintain a set of unique MD5s while processing functions, handling vulnerable functions before non-vulnerable ones. If a function's MD5 already exists in this set, we exclude it from the dataset.

To enhance the realism of our experimental setup, we implement a \textit{time-aware} setting: we use June 1, 2023, as the cutoff date to split training and testing data. 
VFCs submitted after this date comprise the test set, while those submitted before are used for training. We further sort the commits in the training data by submission date. The most recent 10\% of these commits form the validation set, while the earlier 90\% constitute the training set.
Table~\ref{tab:dataset} shows the final data statistics under the time-aware setting.


\begin{table}[t]
\centering
\caption{LMs used in this study.}
\label{tab:model}
\begin{tabular}{@{}cccc@{}}
\toprule
Arch.  & Methods    &  Model   & Parameters \\ 
\midrule
\multirow{5}{*}{Decoder} &  \multirow{5}{*}{\begin{tabular}[c]{@{}c@{}}ICL \\ Fine-tuning\end{tabular}} 
  & \codeqwen  & 6.9B \\
& & \deepseek  & 6.6B \\
& & \codegemma & 8.5B \\
& & \starcoder & 7.2B \\
& & \codellama & 6.6B \\
\midrule
\multirow{3}{*}{Encoder} &  \multirow{3}{*}{Fine-tuning} 
  & CodeBERT       & 125M \\
& & GraphCodeBERT  & 125M \\
& & UniXcoder      & 126M \\
\midrule
\multirow{2}{*}{Enc-Dec} &  \multirow{2}{*}{Fine-tuning} 
  & CodeT5         & 223M \\
& & CodeT5+        & 223M \\ 
\bottomrule
\end{tabular}
\end{table}

\subsection{LMs Selection}
In this work, we select 10 open-source models with varied sizes, architectures, and domains of training data to systematically evaluate their effectiveness on the \task task. 
Following Big Code Models Leaderboard\footnote{https://huggingface.co/spaces/bigcode/bigcode-models-leaderboard} (as of Oct 1, 2024), which evaluates Code LLMs on two common benchmarks, HumanEval~\cite{chen2021codex} and MultiPL-E~\cite{10103177}, We select the five best-performing open-source LLMs including \codeqwen~\cite{qwen}, \deepseek~\cite{guo2024deepseek}, \codegemma~\cite{codegemma_2024}, \starcoder~\cite{lozhkov2024starcoder}, and \codellama~\cite{roziere2023code}, all with versions of approximately 7 billion parameters (7B). 
Therefore, we use the 7B versions in our experiment to ensure a fair comparison.
Where applicable, we use the base versions of the \llms for further classification-based fine-tuning and the chat or instruction-tuned versions for in-context learning.
Additionally, we consider a series of \slms, such as CodeBERT~\cite{feng-etal-2020-codebert}, GraphCodeBERT~\cite{guo2021graphcodebert}, UniXcoder~\cite{guo2022unixcoder}, CodeT5~\cite{wang2021codet5}, and CodeT5+~\cite{wang2023codet5}, which proven effectiveness in a wide range of tasks in the software engineering domain~\cite{hou2023large, 10.1109/ICSE48619.2023.00180}. 
Table~\ref{tab:model} shows additional information on these models, including the architecture, parameters, and how we use them.

Notably, we decided not to include commercial LLMs, e.g., GPT-4 and Claude, for the following three reasons: (1) In certain cases, such as enterprise use, source code may be considered sensitive; sharing those data to \llm service providers may raise issues of data privacy; 
(2) commercial LLMs are not open-source and are updated irregularly, making it difficult to reproduce results from historical versions, which challenges comprehensive and reproducible effectiveness assessments. 
In contrast, open-source LLMs usually provide complete checkpoints as well as clear version records, enabling reproducibility across different model versions and runs; 
(3) many commercial models do not support fine-tuning of model parameters (e.g., Claude), which limits their flexibility for evaluation.

\section{Experimental Design}
\label{sec:setup}

\subsection{RQ1: Effectiveness of LLMs in SVD}
We employ two sets of typical adaptation techniques for \llms in the SE community: prompt-engineering and fine-tuning~\cite{hou2023large, wang2024software, fan2023large}.
Notably, the main difference between prompt-engineering and fine-tuning is whether the model's parameters are updated. 
Prompt engineering operates during model inference and does not involve parameter updates. In this context, the desired behavior of \llm is achieved through carefully structuring the input prompt and incorporating external information into the prompt~\cite{brown2020language}. 
On the other hand, fine-tuning involves updating the model's parameters using a training dataset, allowing the LLM to better adapt to the SVD task.

\subsubsection{Prompt-engineering} 
This approach uses structured prompts to instruct LLMs in identifying whether the provided function code contains vulnerabilities. 
The prompt and the corresponding function code are fed into the LLM, which generates a response containing the classification label. 
The label is subsequently extracted from the generated response.

\begin{table}
    \centering
    \caption{Prompt templates for LLMs.}
    \label{tab:prompt}
    \begin{tabular}{p{0.8\columnwidth}}
        \toprule
        \textbf{Zero-shot Prompt Template} \\
        \midrule
        \textit{\#\#\# Instruction: Analyze the input function, determine if it is vulnerable, and return the answer as the corresponding label ``vulnerable'' or ``non-vulnerable''.} \\
        \textit{\#\#\# Input:} \\
        \textit{\{function\}} \\
        \textit{\#\#\# Response:} \\
        \midrule
        \textbf{In-context \& RAG Prompt Template (1 shot)} \\
        \midrule
        \textit{\#\#\# Instruction: Analyze the input function, determine if it is vulnerable, and return the answer as the corresponding label ``vulnerable'' or ``non-vulnerable''.} \\
        \\
        \textit{\#\#\# Input:} \\
        \textit{\{example function\}} \\
        \textit{\#\#\# Response:} \\
        \textit{\{example label\}} \\
        \\
        \textit{\#\#\# Input:} \\
        \textit{\{function\}} \\        
        \textit{\#\#\# Response:} \\        

        \bottomrule 
    \end{tabular}
\end{table}

Specifically, we explore several types of prompting strategies, including zero-shot generation, In-Context Learning (ICL), and Retrieval-Augmented Generation (RAG)~\cite{hou2023large}.

\begin{itemize}[leftmargin=*]
    \item \textbf{Zero-shot.} 
    The zero-shot prompt starts with a task instruction guiding \llms to analyze the input function and determine if it is vulnerable, followed by the input function code and the response section. The prompt template for zero-shot prompting is outlined in Table~\ref{tab:prompt}.
    
    \item \textbf{ICL.} 
    Different from zero-shot prompting, in the inference stage of ICL, we include $n$ randomly selected examples of function code and corresponding ground-truth labels from training data into the input prompt~\cite{brown2020language}.  
    Specifically, those examples are concatenated after the original task instruction.
    Those in-context examples help LLM better understand the task and align its behavior to match the expected output format~\cite{brown2020language}. 
    
    \item \textbf{RAG.} RAG builds upon the ICL paradigm by incorporating a retrieval mechanism~\cite{lewis2020retrieval}.
    It retrieves relevant information of the given input from a knowledge base and guides \llms to create a contextually relevant response based on the retrieved data. 
    In our approach, we adapt RAG by retrieving the top $n$ example codes most similar to the input function. 
    The similarity between examples is calculated using cosine similarity on code embeddings.     
    We apply SimCSE, a widely used approach for generating code embedding~\cite{gao2021simcse}.
    All the examples are extracted from the training dataset and those examples are concatenated in the input prompt the same way as ICL.
\end{itemize}

\subsubsection{Fine-tuning LLMs} 
Within this category, we adapt \llms to \task task through supervised fine-tuning.
We apply PEFT techniques to fine-tune \llms with vulnerability data~\cite{ding2023parameter}.
Specifically, we use quantized low-rank adaptation (QLoRA), which is a commonly used and lightweight training technique for \llms that has proven to be effective for code-related tasks~\cite{weyssow2023exploring,  saberi2024utilization, silva2023repairllama, esmaeili2024empirical, weyssow2024codeultrafeedback}.
It significantly reduces the number of training parameters of \llms while maintaining the model's performance.

In this study, we investigate two distinct fine-tuning strategies for \llms: (1) instruction tuning, where the training objective is next-token prediction, and (2) sequence classification fine-tuning, where we incorporate a classification head on top of the model backbone and fine-tune \llms with a binary classification loss.

\begin{itemize}[leftmargin=*]
    \item \textbf{Instruction tuning.} 
    In the instruction tuning process, we provide the model with the task instruction, target function, and ground-truth label.  
    We use training set for fine-tuning.
    The model is fine-tuned by predicting the tokens of task instruction, target function, and its label. 
    Following standard practice~\cite{zhang2023instruction}, we minimize the cross-entropy loss between predicted and ground-truth tokens during the tuning process.
    We refer to this loss as causal language modeling (CLM) loss.
    More formally, for each sample $(x_i, y_i)$, where $x_i$ is the instruction with the target function and $y_i$ its ground-truth label, the objective is to minimize the cross-entropy loss:
    \begin{equation}
    \label{eq:loss_inst}
        \mathcal{L}_{CLM} = -\frac{1}{N}\sum_{i=1}^{N} \log P(x_i y_i | \theta)\:,
    \end{equation}
    where $P(x_i y_i | \theta)$ is the probability of generating the tokens $x_i y_i$ parameterized by the model's parameters $\theta$.

    \item \textbf{Sequence classification fine-tuning.} 
    Inspired by Ding et al.~\cite{ding2024vulnerability}, we also fine-tune \llms as a sequence classifier.
    We add a learnable classification head to the \llms, which is a linear layer followed by a softmax function. 
    This layer takes the hidden representation of the last token as input and outputs the classification label.
    The final hidden state contains contextual information from all the preceding tokens.
    In the fine-tuning process, the target function is fed into the \llms, and the model is fine-tuned with binary cross-entropy loss:
    \begin{equation}
    \label{eq:loss_cls}
        \mathcal{L}_{CLS} = -\sum_{i=1}^{N} y_i \log(\hat{y}_i)\:,
    \end{equation}
\end{itemize}
   
    where $N$ is the number of classes,
    $y_i$ is the ground-truth label, and $\hat{y}_i$ is the predicted probability of the $i$-th class.


With instruction tuning, the model is prompted to generate answer text during inference, and the label is extracted via text matching. In contrast, sequence classification fine-tuning involves directly inputting the function into the model during inference, where the model classifies the input as either vulnerable or non-vulnerable.

\subsection{RQ2: Comparison with SLMs and SAST tools}
\noindent To further understand the effectiveness of LLMs in SVD, we also compare their effectiveness with SLMs and SAST tools.

\vspace{4px}
\noindent{\textbf{SLMs.}}
For all the SLMs, following their common use in literature~\cite{niu2023empirical,wang2021codet5,guo2022unixcoder}, we conduct full-parameter fine-tuning, which is also sequence classification fine-tuning.
We first choose CodeBERT, which is the first pre-trained Transformer model on source code.
Based on the recent empirical study comparing multiple SLMs on different code-related tasks~\cite{niu2023empirical}, we further selected CodeT5~\cite{wang2021codet5} and UniXcoder~\cite{guo2022unixcoder}.
CodeT5 and UniXcoder demonstrate superior effectiveness on code understanding tasks.
UniXcoder is a unified cross-modal pre-trained model that leverages multimodal data (i.e., code comment and AST) to pretrain code representation.
We are also aware that CodeT5 has a more recent version, i.e., CodeT5+~\cite{wang2023codet5}, which is enhanced with the flexibility to operate in various modes for different downstream tasks through a mixture of pre-training objectives, which are performed on two stages of pre-training on unimodal and bimodal data.
Consequently, we also include CodeT5+ with the same parameter count as CodeT5. 
In total, we have selected four SLMs, i.e., CodeBERT, UniXcoder, CodeT5, and CodeT5+.

\vspace{4px}
\noindent{\textbf{SAST Tools.}} 
Several studies assessing SAST tools have already been conducted \cite{li2023comparison, kang2022detecting, thung2012extent}. 
Li et al. \cite{li2023comparison} reveal that while SAST tools use rule-based static analysis to detect known vulnerabilities, they may miss more complex or novel issues. In contrast, LLMs are well-known for their abilities to generalize from vast amounts of code and natural language training data \cite{carlini2021extracting}, allowing them to capture context, understand patterns \cite{chen2021codex}, and might detect vulnerabilities that rule-based SAST tools overlook. Given these distinct strengths, it would be interesting to explore and compare the effectiveness of LLMs versus SAST tools in this scenario. 

From all the tools evaluated by the recent empirical study on comparing Java SAST tools~\cite{li2023comparison}, we selected two most popular tools, i.e., Semgrep~\cite{Semgrep:online} and SonarQube~\cite{SonarQube:online}.
These two tools are chosen because they are the most-starred on GitHub among the tools evaluated in the study~\cite{li2023comparison} when writing the paper.
Semgrep had accumulated 9.9k stars~\cite{Semgrep:online}, and the SonarQube community edition (SonarQube) had attracted 8.7k stars~\cite{SonarQube:online} and natively supports multiple programming languages.


Semgrep OSS primarily performs intraprocedural analysis, meaning it analyzes code within individual functions or methods without tracking behavior across function boundaries~\cite{Semgrep:detail}. 
To analyze each VFC in our benchmark, we first identify the files that were modified in that VFC. For each modified file, we retrieve the corresponding file from the parent commit, which contains the vulnerable code. These parent commit files are then input into Semgrep for analysis, which outputs the line numbers of the detected vulnerable code.
We then use Tree-sitter~\cite{treesitter:online} to extract the target functions associated with these line numbers. 
A Semgrep detection is considered a true positive if the target function matches a ground-truth vulnerable function in our benchmark, and a false positive if it does not.
Also, for each of the reported buggy line numbers, Semgrep also outputs the corresponding vulnerable type, but we decided to ignore this information when determining if a Semgrep output is true.

SonarQube performs both intraprocedural and interprocedural analysis, allowing it to detect risks within individual methods as well as across functions and modules. 
Since SonarQube requires entire projects as input, we construct the input projects from the parent commits of each VFC covered in our benchmark.
Once the project is successfully built, SonarQube scans for potential vulnerabilities, generating security issues and hotspots that include the line numbers of identified vulnerabilities. 
We then apply the same evaluation approach used with Semgrep to assess SonarQube’s performance.

\subsection{RQ3: Attempts to improve the effectiveness of LLMs}
Reflecting on the results and data statistics, we aim to develop two strategies, i.e., one from the data perspective and the other from the model perspective, to help improve the model's effectiveness.
From the \textit{data} perspective, one issue is that the training dataset is highly imbalanced. Thus, it is challenging for the model to fully learn to distinguish between vulnerable and non-vulnerable functions.
Therefore, we downsample the non-vulnerable functions to make a balanced dataset, i.e., we have an equal number of vulnerable and non-vulnerable samples in each dataset.
We keep the test set unchanged.

From the \textit{model} perspective, we aim to implement model ensembling, which has proven effective in various software engineering tasks~\cite{xia2015elblocker, kumar2023software}, though its potential benefits remain unexplored for SVD.
Specifically, we implement majority voting with predictions from top-performing LLMs and SLMs. We select the five best-performing models based on validation set results.
We exhaustively evaluate all possible model combinations on the validation set, retaining the combination that yields the highest F1 score. For each test set data point, the final prediction is determined as follows: if more than half of the $k$ models (specifically, at least $\lfloor k/2 \rfloor + 1$ models) classify it as vulnerable, the final label is \texttt{vulnerable}. Otherwise, the label is \texttt{non-vulnerable}.

\subsection{Implementation}
To fine-tune LLMs, we implement parameter-efficient fine-tuning (PEFT) with QLoRA~\cite{dettmers2024qlora} using the HuggingFace library~\cite{wolf-etal-2020-transformers}.
QLoRA reduces the memory usage of LLM fine-tuning without performance tradeoffs compared to standard 16-bit model fine-tuning.
In particular, QLoRA uses 4-bit quantization to compress a pre-trained LLM. 
The LLM parameters are then frozen, and a relatively small number of trainable parameters are added to the model as low-rank adapters. 
During fine-tuning, QLoRA backpropagates gradients through the frozen 4-bit quantized pretrained LLMs into the low-rank adapters. 
The layers of low-rank adapters are the only parameters being updated during fine-tuning.

We apply the recommended hyperparameter settings from Hugging Face’s Trainer.\footnote{\url{https://huggingface.co/docs/transformers/main_classes/trainer}} We use the AdamW optimizer with a learning rate of 5e-5. The batch size is set to 128 for LLMs and 32 for SLMs. We set the rank parameter in LoRA to 16 and alpha to 32.
The whole set of hyper-parameters and the values we used can be found in our replication package.\footnote{\url{https://github.com/soarsmu/SVD-Bench}}

\subsection{Evaluation Metrics}
We formalize the SVD task as a binary classification problem.
The input is a function, and the output should be a binary label, either vulnerable or non-vulnerable.
Following prior works~\cite{li2018vuldeepecker,fu2022linevul,li2021vulnerability}, we use the precision, recall, and F1 score as our evaluation metric. 
As mentioned in Li et al~\cite{li2018vuldeepecker}, it would be fair to say that SVD methods with high false negative rates may not be \textit{useful}, i.e., we want high \textit{recall}. In contrast, those with high false positive rates may not be \textit{usable}, i.e., we want high \textit{precision}.
Thus, we use the F1 score as the main evaluation metric, considering the balance of precision and recall.

\renewcommand{\arraystretch}{1.1}
\setlength{\arrayrulewidth}{.5pt}
\setlength\extrarowheight{.5pt}
\setlength\dashlinedash{2pt}
\setlength\dashlinegap{2pt}
\begin{table*}[!t]
\centering
\footnotesize
\caption{RQ1 Results: the effectiveness of LLMs.} 
\label{tab:rq1_merge}
\resizebox{\textwidth}{!}{%
\begin{tabular}{ll*{9}{>{\centering\arraybackslash}p{1.2cm}}*{9}{c}}
    \toprule
    & & \multicolumn{3}{c}{\textsc{Python}} & \multicolumn{3}{c}{\textsc{Java}} & \multicolumn{3}{c}{\textsc{JavaScript}} \\
    \cmidrule(l{1em}r{1em}){3-5} \cmidrule(l{1em}r{1em}){6-8} \cmidrule(l{1em}r{1em}){9-11}
    \textbf{Model} & \textbf{Method} & \textbf{Precision} & \textbf{Recall} & \textbf{F1} & \textbf{Precision} & \textbf{Recall} & \textbf{F1} & \textbf{Precision} & \textbf{Recall} & \textbf{F1} \\
    \midrule
    \arrayrulecolor{black!50}
    
    \multirow{5}{*}{CodeQwen1.5} & Zero-shot & 0.088 & 0.733 & 0.158 & 0.073 & 0.841 & 0.134 & 0.188 & 0.897 & 0.311 \\
    & ICL & 0.120 & 0.582 & \cellcolor{blue!25} \textbf{0.200} & 0.134 & 0.594 & \cellcolor{blue!25} \textbf{0.219} & 0.164 & 0.440 & 0.239 \\
    & RAG & 0.110 & 0.523 & 0.181 & 0.110 & 0.484 & 0.18 & 0.191 & 0.564 & 0.285 \\
    \cdashline{2-11}
    & QLoRA$_{\text{inst}}$ & 0.0 & 0.0 & 0.0 & 0.0 & 0.0 & 0.0 & 0.310 & 0.606 & 0.410 \\
    & QLoRA$_{\text{cls}}$ & 0.250 & 0.005 & 0.011 & 0.284 & 0.047 & 0.080 & 0.348 & 0.608 & \cellcolor{blue!25} \textbf{0.443} \\

    \midrule

    \multirow{5}{*}{DeepSeek-Coder} & Zero-shot & 0.101 & 0.118 & 0.109 & 0.112 & 0.209 & 0.146 & 0.127 & 0.029 & 0.047 \\
    & ICL & 0.124 & 0.350 & \cellcolor{blue!25} \textbf{0.183} & 0.158 & 0.391 & \cellcolor{blue!25} \textbf{0.225} & 0.149 & 0.231 & 0.181 \\
    & RAG & 0.109 & 0.315 & 0.162 & 0.136 & 0.307 & 0.189 & 0.188 & 0.384 & 0.252 \\
    \cdashline{2-11}
    & QLoRA$_{\text{inst}}$ & 0.0 & 0.0 & 0.0 & 0.103 & 0.007 & 0.014 & 0.209 & 0.708 & 0.323 \\
    & QLoRA$_{\text{cls}}$ & 0.196 & 0.009 & 0.018 & 0.293 & 0.032 & 0.057 & 0.310 & 0.576 & \cellcolor{blue!25} \textbf{0.403} \\

    \midrule

    \multirow{5}{*}{CodeGemma} & Zero-shot & 0.089 & 0.912 & 0.162 & 0.081 & 0.869 & 0.148 & 0.186 & 0.944 & 0.311 \\
    & ICL & 0.124 & 0.263 & \cellcolor{blue!25} \textbf{0.169} & 0.143 & 0.441 & \cellcolor{blue!25} \textbf{0.216} & 0.183 & 0.535 & 0.273 \\
    & RAG & 0.105 & 0.263 & 0.150 & 0.125 & 0.406 & 0.192 & 0.182 & 0.542 & 0.273 \\
    \cdashline{2-11}
    & QLoRA$_{\text{inst}}$ & 0.0 & 0.0 & 0.0 & 0.0 & 0.0 & 0.0 & 0.332 & 0.317 & 0.324 \\
    & QLoRA$_{\text{cls}}$ & 0.308 & 0.014 & 0.026 & 0.333 & 0.073 & 0.120 & 0.345 & 0.514 & \cellcolor{blue!25} \textbf{0.413} \\

    \midrule

    \multirow{5}{*}{StarCoder-2} & Zero-shot & 0.082 & 0.754 & 0.148 & 0.071 & 0.858 & 0.131 & 0.171 & 0.928 & 0.289 \\
    & ICL & 0.104 & 0.033 & 0.050 & 0.118 & 0.114 & 0.116 & 0.179 & 0.297 & 0.224 \\
    & RAG & 0.120 & 0.286 & \cellcolor{blue!25} \textbf{0.169} & 0.163 & 0.193 & \cellcolor{blue!25} \textbf{0.177} & 0.214 & 0.403 & 0.280 \\
    \cdashline{2-11}
    & QLoRA$_{\text{inst}}$ & 0.0 & 0.0 & 0.0 & 0.0 & 0.0 & 0.0 & 0.336 & 0.617 & \cellcolor{blue!25} \textbf{0.435} \\
    & QLoRA$_{\text{cls}}$ & 0.333 & 0.000 & 0.001 & 0.293 & 0.041 & 0.072 & 0.338 & 0.582 & 0.428 \\

    \midrule

    \multirow{5}{*}{CodeLlama} & Zero-shot & 0.083 & 0.619 & 0.147 & 0.069 & 0.727 & 0.126 & 0.178 & 0.955 & 0.299 \\
    & ICL & 0.105 & 0.202 & 0.138 & 0.102 & 0.383 & 0.161 & 0.181 & 0.592 & 0.278 \\
    & RAG & 0.120 & 0.286 & \cellcolor{blue!25} \textbf{0.169} & 116 & 0.447 & \cellcolor{blue!25} \textbf{0.184} & 0.193 & 0.661 & 0.299 \\
    \cdashline{2-11}
    & QLoRA$_{\text{inst}}$ & 0.0 & 0.0 & 0.0 & 0.0 & 0.0 & 0.0 & 0.277 & 0.635 & \cellcolor{blue!25} \textbf{0.386} \\
    & QLoRA$_{\text{cls}}$ & 0.159 & 0.003 & 0.007 & 0.296 & 0.060 & 0.100 & 0.270 & 0.590 & 0.370 \\

    \arrayrulecolor{black}
    \bottomrule
\end{tabular}
}
\begin{flushleft}
\small
\textit{Note:  QLoRA$_{\text{cls}}$ refers to LLM with sequence classification fine-tuning, QLoRA$_{\text{inst}}$ refers to LLM with instruction tuning.} 
\end{flushleft}

\end{table*}

\section{Results}
\label{sec:results}

\subsection{RQ1: Investigating the effectiveness of LLMs.} 

\subsubsection{Effectiveness across \llms}
Table~\ref{tab:rq1_merge} presents the results of applying prompt engineering and fine-tuning techniques to LLMs across three PLs. 
Firstly, we observe the effectiveness of LLMs varies across PLs. 
In Python and Java, the results for all \llms are unsatisfactory, with the best LLM (i.e., CodeQwen1.5 and DeepSeek-Coder) achieving F1 scores of only 0.200 and 0.225, respectively. 
In contrast, LLMs consistently perform better on the JavaScript dataset, with the instruction-tuned StarCoder-2 achieving the highest F1 score of 0.435.
While there is no single LLM that consistently outperforms others across all strategies, CodeQwen1.5, with proper adaptation strategy, achieves the best effectiveness in Python and Java and comparable effectiveness to the best performer in the JavaScript dataset. 
Based on these findings, we recommend that future research consider CodeQwen1.5 as an initial point for further exploration.

\subsubsection{Effectiveness of prompt-engineering}
Table~\ref{tab:rq1_merge} also demonstrates that prompt engineering generally enhances the effectiveness of \llms in Python and Java. 
Specifically, four out of five LLMs using ICL prompting show improved F1 scores in Python, with increases ranging from 4.32\% to 67.89\%, while three show improvements in Java, ranging from 27.78\% to 54.11\%.
Similarly, RAG results in F1 score improvements for four out of five LLMs compared to zero-shot prompting in Python, and for all LLMs in Java.
However, prompt-engineering approaches do not yield improvements for JavaScript. Specifically, the effectiveness of four out of five \llms decreases by up to 23.15\% in terms of F1 score with prompt engineering. 
We hypothesize that this effectiveness decline may be due to LLMs being trained with a greater proportion of JavaScript data, given that it has been the most popular PL on GitHub over the past decade.\footnote{\url{https://github.blog/news-insights/research/the-state-of-open-source-and-ai/\#the-most-popular-programming-languages}} As a result, the additional knowledge injected through prompt-engineering may be less effective and could introduce noise, thereby reducing effectiveness.


\subsubsection{Effectiveness of fine-tuning}
We also observe that fine-tuning substantially improves \llms effectiveness in the JavaScript data but is ineffective in Python and Java.
The effectiveness of all \llms after fine-tuning is lower than that of vanilla \llms with zero-shot prompting in Python and Java. 
Specifically, the F1 score for all LLMs, except \deepseek, drops to 0 following instruction tuning in Python and Java. 
Additionally, LLMs with sequence classification fine-tuning exhibit substantial effectiveness declines, with F1 scores decreasing by 83.95\% to 99.32\% for the Python data and 18.92\% to 60.96\% for the Java data.
In contrast, fine-tuning boosts LLM effectiveness in JavaScript. Compared to zero-shot prompting \llms, instruction tuning increases the F1 score for JavaScript by 4.18\% to 587.23\%, and sequence classification fine-tuning by 32.80\% to 757.4\%, respectively.
We hypothesize that these discrepancies across PLs may be attributed to varying degrees of training data imbalance. Specifically, the ratio of vulnerable to non-vulnerable functions is 1:2 in the JavaScript training dataset, compared to 1:16 in Python and 1:10 in Java. 
We will further investigate and explore the impact of data imbalance on model effectiveness in RQ3.

\subsubsection{Error analysis of \llms}
Given the unsatisfactory performance of LLMs, which falls significantly short of practical application, conducting a detailed error analysis is essential. 
The deeper examination of LLM's predictions allows for a more nuanced understanding of the model's strengths and limitations in addressing the \task task.
In particular, we analyze the distribution of Common Weakness Enumeration (CWE) types in LLM predictions. 
By investigating which CWE types LLMs are most and least effective at identifying, we can infer LLM’s strengths and weaknesses in detecting specific vulnerabilities.
For analysis, we select the best-performing LLM for each PL.

Firstly, we extract the number of CWE types in the test data that LLMs correctly identify, compared to the ground-truth CWE distribution in the test dataset.
The results are presented in Table~\ref{tab:rq1_cwe_distribution}. 
Notably, we observe that the top-performing LLMs for Python and Java, adapted using ICL prompting without fine-tuning, cover 93.75\% and 89.19\% CWE types in their respective test sets. 
This indicates that LLMs may possess prior knowledge of vulnerabilities across various CWE types in Python and Java to some extent.

As the best-performing LLM for JavaScript, i.e., CodeQwen1.5 with sequence classification fine-tuning, covers only 30.23\% of the CWE types in the test set, we further investigate whether this performance gap stems from the fine-tuning process. 
Specifically, we compare it to the best-performing LLM using prompt-engineering approaches for JavaScript: CodeGemma with zero-shot prompting. 
Notably, the correct predictions by CodeGemma with zero-shot prompting cover all CWE types in the test set, supporting our hypothesis that LLMs may retain prior knowledge of vulnerabilities across diverse CWE types and that the fine-tuning process may alter this inherent knowledge.

\begin{figure}
    \centering
    \includegraphics[width=\linewidth]{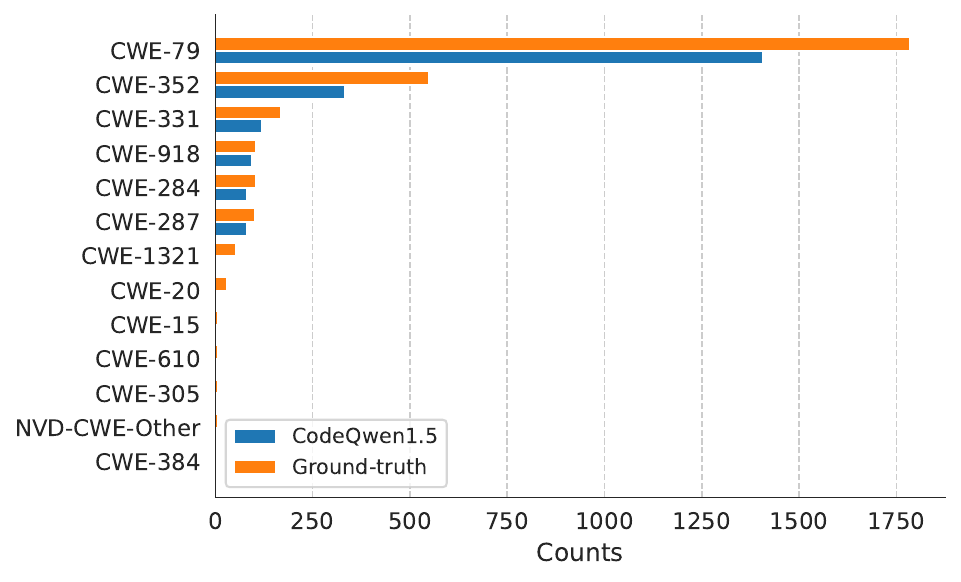}
    \caption{Number of instances for each CWE category correctly predicted by CodeQwen1.5. CWE categories with no correctly identified test instances by CodeQwen1.5 are omitted.}
    \label{fig:cwe_distribution}
\end{figure}

We further analyze how fine-tuning alters the inherent knowledge of LLMs.
Our findings reveal a correlation between the distribution of CWE types in the training data and the model's post-fine-tuning performance in identifying corresponding vulnerabilities within the JavaScript dataset.
Figure~\ref{fig:cwe_distribution} illustrates this relationship by comparing the number of instances correctly identified by CodeQwen1.5 after sequence classification fine-tuning against the ground truth in the test dataset.
Remarkably, CodeQwen1.5 demonstrates exceptional effectiveness, achieving over 60\% accuracy in identifying vulnerable functions across seven CWE types.
Of these seven, six are among the 20 most prevalent CWE types in the training dataset.
Conversely, CodeQwen1.5 fails to identify any vulnerable functions for 30 CWE types, 21 of which each represent less than 1\% of the vulnerable functions in the training dataset, highlighting a strong correlation between data representation and model proficiency.
This correlation underscores how the distribution of CWE types in the training data impacts the model's effectiveness in identifying vulnerabilities for corresponding CWE types.
Addressing this imbalance will be a key focus in our future research to enhance the model's ability to detect a broader range of vulnerabilities consistently.

\vspace{4px}
\noindent\fbox{\begin{minipage}{12.8cm}
\textbf{Answer to RQ1:} \textit{The performance of LLMs varies across PLs. In general, prompt-engineering approaches enhance LLM performance in Python and Java, while fine-tuning proves more beneficial for LLMs in JavaScript.
}
\end{minipage}}


\begin{table}[]
\caption{CWE distribution for LLM's correct prediction in the test set.} 
\label{tab:rq1_cwe_distribution}
\begin{tabular}{@{}lllcc@{}}
\toprule
 & Strategy & PL & \# CWE$_{\text{covered}}$ & \# CWE$_{\text{overall}}$ \\ \midrule
CodeQwen1.5 &   ICL  & Python    & 75         & 80                  \\
DeepSeek-Coder & ICL  & Java  & 33                  & 37                  \\
CodeQwen1.5 & QLoRA$_{\text{cls}}$ & JavaScript    &  13                 & 43                  \\ \bottomrule
\end{tabular}
\end{table}

\subsection{RQ2: Comparison with SLMs and SAST Tools}

\begin{figure}
    \centering
    \includegraphics[width=0.7\linewidth]{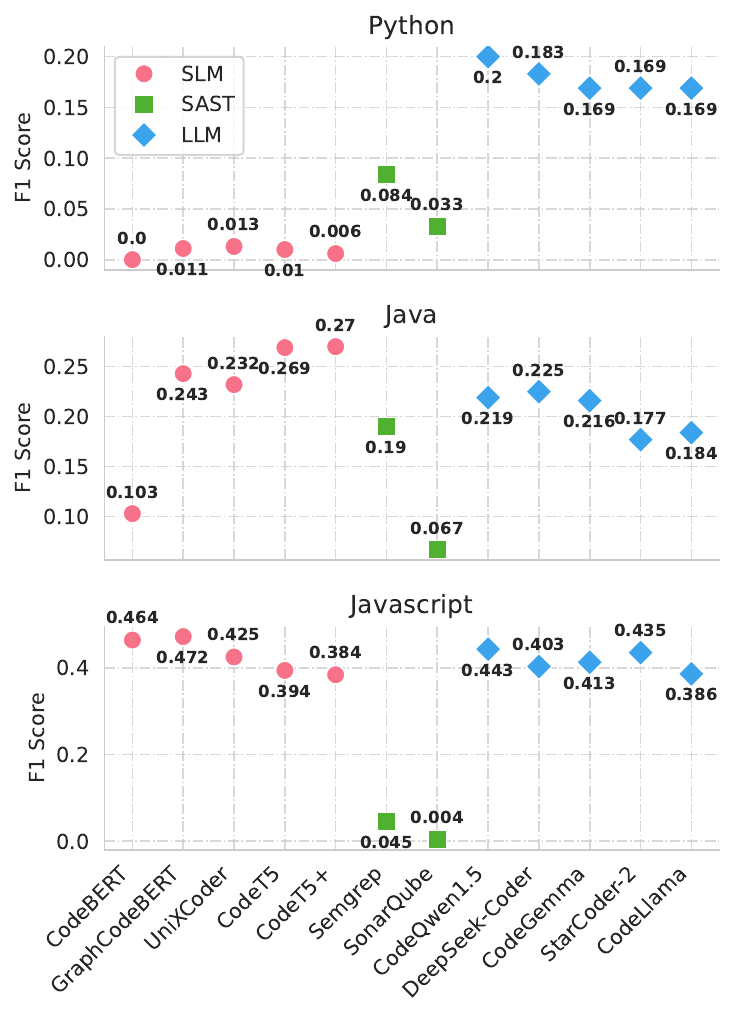}
    \caption{RQ2 Results: Comparison of LLMs with SLMs and SAST tools.}
    \label{fig:rq2_results}
\end{figure}

Figure~\ref{fig:rq2_results} shows the results from SLMs and SAST tools.
We observe effectiveness variations across PLs.
Notably, all of the SLMs underperform each LLM on the zero-shot setting in Python.
However, SLMs outperform LLMs in both Java and JavaScript.
The best-performing SLM in Java, CodeT5+, surpasses the best LLM strategy, i.e., \deepseek with ICL prompting, by 20.00\% in terms of F1. 
Similarly, GraphCodeBERT performs best in JavaScript, exceeding the top LLM strategy, i.e., CodeQwen1.5 with sequence classification fine-tuning, by 4.74\% in terms of F1.
Furthermore, we observe that the effectiveness of SAST tools is consistently lower than LLMs among all PLs, suggesting the great potential for LMs for \task.

\vspace{4px}
\noindent\fbox{\begin{minipage}{12.8cm}
\textbf{Answer to RQ2:} 
\textit{SLMs outperform LLMs in Java and JavaScript but underperform compared to LLMs in Python. SAST tools consistently perform worse than LLMs and SLMs across all PLs.}
\end{minipage}}

\subsection{RQ3: Improving the effectiveness of LLMs}
\subsubsection{Data Perspective}

Table~\ref{tab:rq3_result_downsampling} presents the results of fine-tuning LLMs with downsampled data. For Python and Java datasets, models trained on balanced data achieved higher F1 scores on the same test data. 
This improvement trend is evident in both instruction-tuning and sequence classification fine-tuning. 
However, the JavaScript dataset showed mixed effects when trained on balanced data. In instruction tuning, while three models achieved higher F1 scores, two models produced lower scores. 
Similarly, in sequence classification fine-tuning, four out of five models show decreased F1 scores when trained on balanced data. The raw JavaScript dataset differs notably from the other two PLs as it is inherently more balanced with a 1:2 ratio of vulnerable to non-vulnerable functions. This initial balance may explain why further balancing did not consistently improve effectiveness. Future research could explore how varying ratios of vulnerable to non-vulnerable functions affect model effectiveness.

\begin{table}[h]    
    \centering
    \caption{RQ3 Results: data downsampling.}
    \begin{tabular}{cclrrr}
    \toprule
    \textbf{PL} & \textbf{Method} & \textbf{Model} & \textbf{Precision} & \textbf{Recall} &  \textbf{F1}  \\
     \midrule
    \multirow{10}{*}{Python} & \multirow{5}{*}{Instruction}
       & \codeqwen  & 0.119 & 0.576 & 0.197 $\uparrow$ \\
     & & \deepseek  & 0.103 & 0.668 & 0.178 $\uparrow$ \\
     & & \codegemma & 0.124 & 0.372 & 0.186 $\uparrow$ \\ 
     & & \starcoder & 0.091 & 0.791 & 0.164 $\uparrow$ \\ 
     & & \codellama & 0.119 & 0.600 & 0.198 $\uparrow$ \\ 
     \cmidrule{2-6}
     & \multirow{5}{*}{Classification}
       & \codeqwen  & 0.117 & 0.671 & 0.199  $\uparrow$ \\
     & & \deepseek  & 0.129 & 0.541 & \cellcolor{blue!25} \textbf{0.209}  $\uparrow$ \\
     & & \codegemma & 0.122 & 0.574 & 0.201  $\uparrow$ \\
     & & \starcoder & 0.117 & 0.730 & 0.201  $\uparrow$ \\
     & & \codellama & 0.110 & 0.793 & 0.194  $\uparrow$ \\
    \midrule
    \multirow{10}{*}{Java} & \multirow{5}{*}{Instruction}
       & \codeqwen  & 0.143 & 0.559 & 0.228 $\uparrow$ \\
     & & \deepseek  & 0.132 & 0.634 & 0.218 $\uparrow$ \\
     & & \codegemma & 0.160 & 0.245 & 0.194 $\uparrow$ \\ 
     & & \starcoder & 0.153 & 0.514 & 0.235 $\uparrow$ \\ 
     & & \codellama & 0.152 & 0.484 & 0.232 $\uparrow$ \\ 
     \cmidrule{2-6}
     & \multirow{5}{*}{Classification}
       & \codeqwen  & 0.158 & 0.503 & 0.241 $\uparrow$\\
     & & \deepseek  & 0.159 & 0.591 & 0.251 $\uparrow$\\
     & & \codegemma & 0.185 & 0.331 & 0.237 $\uparrow$\\ 
     & & \starcoder & 0.184 & 0.436 & 0.258 $\uparrow$\\ 
     & & \codellama & 0.175 & 0.572 & \cellcolor{blue!25} \textbf{0.268} $\uparrow$\\ 
    \midrule    
    \multirow{10}{*}{JavaScript} & \multirow{5}{*}{Instruction}
       & \codeqwen  & 0.358 & 0.652 & 0.462 $\uparrow$ \\
     & & \deepseek  & 0.215 & 0.682 & 0.326 $\uparrow$ \\
     & & \codegemma & 0.364 & 0.427 & 0.393 $\uparrow$ \\ 
     & & \starcoder & 0.226 & 0.695 & 0.341 $\downarrow$ \\ 
     & & \codellama & 0.202 & 0.724 & 0.316 $\downarrow$ \\
     \cmidrule{2-6}
     & \multirow{5}{*}{Classification}
       & \codeqwen  & 0.376 & 0.609 & \cellcolor{blue!25} \textbf{0.465} $\uparrow$\\
     & & \deepseek  & 0.259 & 0.620 & 0.365 $\downarrow$ \\
     & & \codegemma & 0.269 & 0.629 & 0.377 $\downarrow$ \\
     & & \starcoder & 0.255 & 0.679 & 0.371 $\downarrow$ \\
     & & \codellama & 0.234 & 0.606 & 0.337 $\downarrow$ \\
     \bottomrule
    \end{tabular}
    \label{tab:rq3_result_downsampling}
    \vspace{1mm} 
    \begin{flushleft}
    \textit{Note:} $\uparrow$ indicates F1 score higher than the model trained with original data; $\downarrow$ indicates F1 score lower than the model trained with original data.
    \end{flushleft}
\end{table}

\begin{figure}[h]
    \centering
    \begin{subfigure}[b]{0.3\textwidth}
        \centering
        \includegraphics[width=\textwidth]{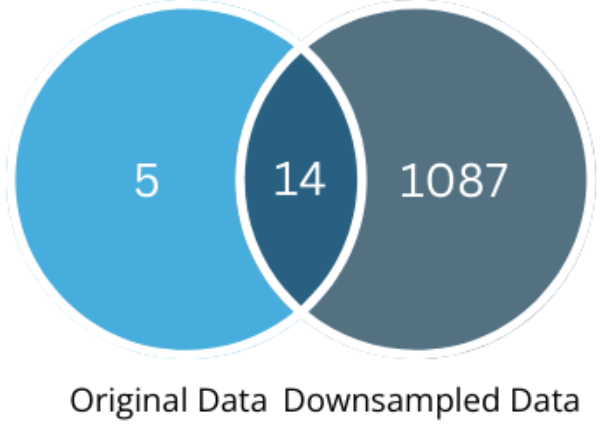}
        \caption{On Python dataset}
        \label{fig:py_downsample}
    \end{subfigure}
    \hspace{20mm}
    \begin{subfigure}[b]{0.3\textwidth}
        \centering
        \includegraphics[width=\textwidth]{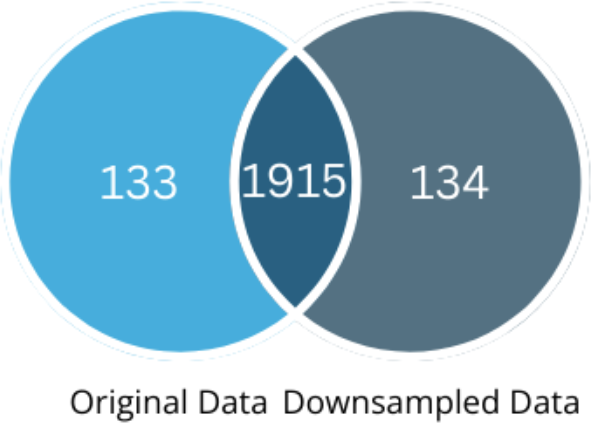}
        \caption{On JavaScript dataset}
        \label{fig:js_downsample}
    \end{subfigure}
    \caption{RQ3: Comparison between fine-tuning with original data and downsampled data.}
    \label{fig:combined_downsample}
\end{figure}

To better understand the differences in results between fine-tuning with original and downsampled data, we created Venn diagrams illustrating correct predictions of vulnerable functions. 
For Python, we selected the best-performing model trained on the downsampled data: \deepseek with classification loss. 
Similarly, for JavaScript, we chose \codeqwen with classification loss. 
Figure~\ref{fig:combined_downsample} demonstrates that for the Python dataset, \deepseek correctly predicts significantly more vulnerable functions when trained on downsampled data. However, we found that 5 unique functions could only be correctly identified using the original training data. This suggests that simply reducing the number of non-vulnerable functions may impair the model's effectiveness in predicting vulnerable functions.

The difference between fine-tuning with original and downsampled data is less pronounced in the JavaScript dataset. There is substantial overlap (87.8\%, or 1,915 out of 2,182) in successful predictions between models trained on original and downsampled data. Nevertheless, models trained on different datasets still have unique correct predictions. This indicates that when fine-tuning with more balanced data, the majority of correctly predicted functions remain consistent.

When we reduced the number of non-vulnerable functions to match the number of vulnerable functions, the model trained on this balanced dataset correctly predicted 134 vulnerable functions while sacrificing 133 correct predictions. This further supports the observation that simply reducing the number of non-vulnerable functions may compromise the model's effectiveness in predicting vulnerable functions.


\subsubsection{Model Perspective}
\begin{table}[ht]
    \centering
    \caption{RQ3 Results: ensemble learning.}
    \begin{tabular}{c c c r r r}
    \toprule
    \textbf{Strategy} & \textbf{PL} & \textbf{Precision} & \textbf{Recall} &  \textbf{F1}  \\
    \midrule
    \multirow{3}{*}{LLMs-only} & Python & 0.083 & 0.619 & 0.147 $\downarrow$\\
     \cmidrule{2-5}
        & Java & 0.186 & 0.338 & 0.240 $\uparrow$ \\
     \cmidrule{2-5}
      & JavaScript & 0.209 & 0.708 & 0.323 $\downarrow$\\
      \midrule
      \multirow{3}{*}{LLMs + SLMs} & Python & 0.083 & 0.619 & 0.147 $\downarrow$\\
     \cmidrule{2-5}
        & Java & 0.186 & 0.338 & 0.240 $\uparrow$\\
     \cmidrule{2-5}
      & JavaScript & 0.209 & 0.708 & 0.323 $\downarrow$\\
      \midrule
      \multirow{3}{*}{LLMs + SAST tools} & Python & 0.083 & 0.619 & 0.147 $\downarrow$\\
     \cmidrule{2-5}
        & Java & 0.186 & 0.338 & 0.240 $\uparrow$\\
     \cmidrule{2-5}
      & JavaScript & 0.209 & 0.708 & 0.323 $\downarrow$\\
    \bottomrule
    \end{tabular}    
    \label{tab:rq3_result_ensemble}
\begin{flushleft}
\small
\textit{Note:} $\uparrow$ indicates F1 score higher than the best F1 we achieve without ensemble learning; $\downarrow$ indicates F1 score lower than the best F1 we achieve without ensemble learning.
\end{flushleft}
\end{table}

Table~\ref{tab:rq3_result_ensemble} presents the results of our ensemble learning approach, considering three scenarios: the first one involving only LLMs, the second one incorporating both LLMs and SLMs, and the third one incorporating both LLMs and SAST tools. 
Contrary to expectations, ensemble learning using LLMs alone does not yield improvements over single-model performance. 
For the Java dataset, ensemble learning achieves an F1-score of 0.24, whereas fine-tuning CodeT5+ as a single model attains a higher F1-score of 0.27. 
Similarly, on the JavaScript dataset, ensemble learning produces an F1-score of 0.323, while fine-tuning GraphCodeBERT as a single model reaches a superior F1-score of 0.472. 
Notably, the results of ensemble learning remain identical when including both LLMs and SLMs or LLMs and SAST tools, indicating that SLMs and SAST tools are not selected in the optimal model combination for the validation set. 
This seemingly contradictory outcome can be attributed to potential differences in data distribution across training, validation, and testing sets. 
Our findings suggest that in a time-aware split setting, selecting models based solely on validation set performance does not guarantee optimal results on the test set. 
Future research should address the domain shift phenomenon prevalent in real-world data.

\vspace{4px}
\noindent\fbox{\begin{minipage}{12.8cm} \textbf{Answer to RQ3:} \emph{Our results show that (1) Fine-tuning LLMs with downsampled balanced data can enhance F1 scores compared to fine-tuning with original data for Python and Java. However, the effect on JavaScript is inconsistent. (2) Simple majority voting proves ineffective in the time-aware split setting.}
\end{minipage}}\\

\section{Discussion}
\label{sec:discussion}

\subsection{Implications}
Reflecting on the experimental results from Section~\ref{sec:results}, we identify the following implications for future research.

\vspace{4px}
\noindent{\textbf{Training data statistics play a vital role in SVD.} 
Based on Table~\ref{tab:rq1_merge}, we can find that the optimal model and corresponding strategy vary across different PLs. 
Considering the different characteristics of the datasets, we infer that few-shot learning tends to excel with highly imbalanced training data while fine-tuning yields better performance when more balanced data is available. Therefore, training data statistics must be carefully considered when selecting an appropriate strategy.}

\vspace{4px}
\noindent{\textbf{High-quantity vulnerability data benefits SVD performance.} 
A comparison of model performance across Python, Java, and JavaScript datasets reveals a significant correlation between data quantity and results. The best-performing model achieves an F1-score of 0.44s with sequence classification fine-tuning on JavaScript, with the largest number of vulnerable functions in the training set. In contrast, Python and Java datasets, containing fewer vulnerable functions, yield F1-scores below 0.2 using fine-tuning. This disparity suggests that increasing the volume of vulnerability data may enhance the capabilities of LLMs in SVD.
}

\vspace{4px}
\noindent{\textbf{Model-specific performance variations require tailored approaches.}
Although the five LLMs are all based on decoder-only architecture, we observe variations in performance across different strategies. 
For instance, \deepseek shows improved results with ICL and RAG compared to zero-shot approaches across all three PL datasets. 
Conversely, \starcoder performs less when additional random shots were introduced as ICL.
These findings underscore the model-specific nature of performance and the varying effectiveness of zero-shot, ICL, and RAG strategies across different tasks. Given this unpredictability, we recommend conducting preliminary experiments on a sample of data to determine the most effective approach -- whether zero-shot or few-shot for each specific model and task combination. This insight highlights the critical importance of tailored strategies in LLM application, emphasizing the need for empirical testing to optimize performance in diverse contexts.}

\subsection{Threats to Validity}
\noindent \textbf{Threats to Construct Validity.} A potential threat to construct validity lies in our choice of evaluation metrics. While we follow established practices in SVD research~\cite{li2018vuldeepecker,fu2022linevul,li2021vulnerability} using precision, recall, and F1-score, these metrics may not always capture the full complexity of real-world scenarios.
In different contexts, the relative importance of precision and recall may vary. For example, when false positives significantly impede developer productivity, greater emphasis might be placed on precision. In such cases, alternative metrics like the F0.3-score (where precision is weighted twice as heavily as recall) could be more appropriate.
Nevertheless, given the widespread use and acceptance of the F1-score in the field, we believe this threat is minimal. Our choice aligns with standard practices, facilitating comparisons with existing literature while providing a balanced assessment of model performance.

\vspace{4px}
\noindent \textbf{Threats to Internal Validity.} 
One challenge arises from the prediction methods employed by Semgrep and SonarQube. As these tools do not generate function-level predictions directly, we must input coarser-grained data (files or entire projects). To maintain consistency in our function-level evaluation, we only consider their predictions pertaining to functions within our test set. While this approach may produce additional false alarms outside our test set, we do not count these as false positives. Based on our current knowledge, we believe this strategy offers the most equitable evaluation, thus minimizing this threat to internal validity.

\vspace{4px}
\noindent{\textbf{Threats to External Validity.}} One potential external threat is we only work on vulnerabilities recorded in the NVD, and the results may not be able to be generalized to other vulnerabilities from other resources.
In the future, we aim to extend our experiments to include vulnerabilities in diverse resources.

\section{Conclusion and Future Work}
\label{sec:conclusion}
In this work, we conduct a comprehensive empirical study evaluating LLMs on the SVD task. We curated a dataset comprising vulnerabilities across three different PLs: Python, Java, and JavaScript. To ensure the dataset's quality, we leveraged the VFC information recorded in the NVD.
To understand how LLMs perform, we first comprehensively implement prompt engineering and fine-tuning (instruction tuning and sequence classification fine-tuning). 
Second, we compare LLMs with SLMs and SAST tools. Third, we attempt to enhance the effectiveness of LLMs by training the models with downsampled data and conducting ensemble learning.

In the context of SVD, our research reveals that fine-tuning is the optimal strategy for adapting LLMs when sufficient and well-balanced training data is available. 
However, in scenarios with limited data, LLMs utilizing few-shot learning techniques prove more effective.
Our experiments to enhance LLM effectiveness demonstrate that training models with downsampled data yield benefits when the original dataset is highly imbalanced. In cases where the data is already well-balanced, the improvement is minimal and may lead to decreased effectiveness. Additionally, we found that simple majority voting does not significantly improve the model's F1 score on the SVD task.
These insights help better understand LLM performance in SVD and lay the groundwork for improving LLM effectiveness in identifying vulnerabilities.

In future work, we are interested in exploring additional methods to improve LLMs for SVD, such as incorporating more advanced prompt engineering techniques. 
We also plan to investigate more sophisticated ensemble learning methods and LLM collaboration strategies to further aggregate successful predictions from different models.

\newpage
\bibliography{main}
\end{document}